\documentclass[aps,prl,twocolumn,showpacs]{revtex4}

\usepackage{graphicx}
\usepackage{dcolumn}
\usepackage{bm}
\usepackage{epstopdf}
\usepackage[applemac]{inputenc}
\usepackage[francais]{babel}
\usepackage[T1]{fontenc}

\newcommand{\bq}{\begin{equation}}
\newcommand{\eq}{\end{equation}}
\newcommand{\bqa}{\begin{eqnarray}}
\newcommand{\eqa}{\end{eqnarray}}
\newcommand{\nn}{\nonumber \\}

\def\be     {\begin{equation}}
\def\ee     {\end{equation}}
\def\bea        {\begin{eqnarray}}
\def\eea        {\end{eqnarray}}
\def\bnn    {\begin{eqnarray*}}
\def\enn    {\end{eqnarray*}}

\begin{document}

\title{The thermopower as a fingerprint of the Kondo breakdown quantum critical point}

\author{K.-S. Kim${}^{1,3}$ }
\author{C. P\'epin${}^{2,4}$}

\affiliation{$^1$Asia Pacific Center for Theoretical Physics,
Hogil Kim Memorial building 5th floor, POSTECH, Hyoja-dong, Namgu,
Pohang 790-784, Korea\\ $^2$Institut de Physique Th\'eorique, CEA,
IPhT, CNRS, URA 2306, F-91191 Gif-sur-Yvette, France\\
$^3$Department of Physics, Pohang University of Science and
Technology, Pohang, Gyeongbuk 790-784, Korea \\ $^4$International
Institute of Physics, Universidade Federal do Rio Grande do Norte,
59078-400 Natal-RN, Brazil }

\date{\today}

\begin{abstract}
We propose that the thermoelectric power distinguishes two
competing scenarios for quantum phase transitions in heavy
fermions : the spin-density-wave (SDW) theory and breakdown of the
Kondo effect.
%
%The thermopower is proposed as a decisive measurement
%distinguishing two competing scenarios for quantum phase
%transitions in heavy fermions :  the spin-density-wave (SDW)
%theory and the Kondo breakdown one.
%
In the Kondo breakdown scenario, the Seebeck coefficient turns out
to collapse from the temperature scale $E^{*}$, associated with
quantum fluctuations of the Fermi surface reconfiguration. This
feature differs radically from the physics of the SDW theory,
where no reconstruction of the Fermi surface occurs, and can be
considered as the hallmark of the Kondo breakdown theory. We test
these ideas, upon experimental results for YbRh$_2$Si$_2$.
\end{abstract}


\maketitle

%The energy transfer from heat to electricity (Seebeck effect) and
%vice versa (Peltier effect) have been ones of the most studied
%problems in the industry \cite{Seebeck_Peltier}.  Comparatively
%the thermopower has been little studied in the context of quantum
%liquids, compared with other transport coefficients such as both
%electrical and thermal conductivities \cite{Mahan_Comment}. One
%recent thermopower measurement \cite{hartmann} in the  heavy
%fermion (HF) compound YbRh$_2$Si$_2$ suggests that it can be an
%important experiment, particularly in order to distinguish between
%the two main scenarios for quantum criticality in these compounds
%\cite{HF_Review}, that is, the spin-density-wave (SDW) theory
%\cite{HMM,Rosch} and the Kondo breakdown
%\cite{KB_Senthil,KB_Indranil,KB_Pepin}.

A quantum transition from a light metal to a heavy Fermi liquid is
believed to occur in a class of heavy fermion (HF) compounds as a
result of the competition between Rudderman-Kittel-Kasuya-Yoshida
interactions and the formation and condensation of Kondo singlets
\cite{HF_Review}. The Kondo breakdown theory
\cite{KB_Senthil,KB_Indranil,KB_Pepin} offers a minimal model
describing this transition. It differs from the spin-density-wave
(SDW) theory \cite{HMM,Rosch} in respect that the whole heavy
Fermi surface is destabilized at the quantum critical point (QCP)
and a reconstruction of the Fermi surface is inevitable, sometimes
identified with an orbital selective Mott transition. As a result,
the Kondo breakdown QCP is described by critical fluctuations with
the dynamical exponent $z = 3$, associated with breakdown of the
Kondo effect \cite{KB_Indranil,KB_Pepin}, while the SDW QCP is
explained within $z = 2$ related with antiferromagnetic spin
fluctuations \cite{HMM,Rosch}.

Until now, the thermopower has been studied mainly in the heavy
Fermi liquid phase, with a special focus on its sign and its ratio
with the Sommerfeld coefficient. In heavy fermions, the sign of
the thermopower is determined by the position of the Kondo
resonance with respect to the Fermi surface. The strong mass
renormalization is proven not to affect the quasi-universal ratio
to the Sommerfeld coefficient, which remains almost the same as in
conventional metals \cite{Sign_Seebeck,HF_Seebeck}. Little is
known about the Seebeck coefficient close to the QCP. Preliminary
studies for CeCu$_{6-x}$Au$_x$ \cite{lohneysen} and
Ce(Ni$_{1-x}$Pd$_x$)$_2$Ge$_2$ \cite{kuwai} show that the presence
of a QCP modifies low temperature dependence of the Seebeck
coefficient. Two recent studies under magnetic fields show some
striking similarity between thermoelectric effects in CeCoIn$_5$
\cite{kamran2} and URu$_2$Si$_2$\cite{kamran3}. In particular,
both systems exhibit a pronounced anisotropy in their
thermoelectric response. Lastly, a recent experiment on
YbRh$_2$Si$_2$ under weak magnetic fields shows some drastic
variations of the magnitude of the Seebeck coefficient in both
sides of the QCP \cite{hartmann}. Even fewer theoretical studies
are available \cite{Sign_Seebeck,pruschke,zlatic}. In the case of
the SDW the authors of Ref. \cite{indranil} argued that the
Seebeck coefficient divided by temperature has the same variation
in temperature as the Sommerfeld coefficient and the
quasi-universal ratio is preserved at the SDW QCP.

In this study we show that the thermoelectric power can be
regarded as one of the hallmarks for the Kondo breakdown quantum
criticality, providing a careful fit to the recent experimental
observations on YbRh$_2$Si$_2$ \cite{hartmann}. Precisely, the
abrupt collapse from the temperature identified with $E^{*}$ in
the Kondo breakdown scenario \cite{KB_Indranil,KB_Pepin} turns out
to be the unique prediction from the Kondo breakdown theory beyond
the SDW \cite{HMM,Rosch} and local quantum critical \cite{Si}
scenarios.

%In this study we want to test the prediction of the minimal model
%for the Kondo breakdown for the Seebeck coefficient
%\cite{KB_Senthil,KB_Indranil,KB_Pepin}. We argue that it is mainly
%sensitive to quantum fluctuations of the Fermi surface
%reconfiguration, a fingerprint of the Kondo breakdown mechanism.
%Precisely, the Seebeck coefficient turns out to collapse from the
%temperature identified with $E^{*}$ in the Kondo breakdown
%scenario \cite{KB_Indranil,KB_Pepin}, the energy scale of which is
%associated with quantum fluctuations of the Fermi surface
%reconfiguration in both the quantum critical regime and
%antiferromagnetic phase. On the other hand, the thermoelectric
%power displays moderate saturation in both the antiferromagnetic
%and paramagnetic Fermi liquids of the SDW theory because the Fermi
%surface reconfiguration does not exist in such a theory. We
%confirm the validity of the minimal model for the Kondo breakdown
%by providing a fit to the recent experimental observations on
%YbRh$_2$Si$_2$. \cite{hartmann}.

%Thermopower is the transport coefficient to measure how much the
%change of voltage is induced by temperature gradient, given by the
%ratio between the electrical conductivity and the coefficient for
%the thermal-gradient induced electric-current. The Seebeck
%coefficient divided by temperature is well known in conventional
%metals described by Fermi liquid, given by a small constant at low
%temperatures owing to particle-hole symmetry while attributed to
%phonons at high temperatures.

We start from the U(1) slave-boson representation of the Anderson
lattice model in the large-$U$ limit \bqa && L_{ALM} = \sum_{i}
c_{i\sigma}^{\dagger}(\partial_{\tau} - \mu)c_{i\sigma} - t
\sum_{\langle ij \rangle} (c_{i\sigma}^{\dagger}c_{j\sigma} +
H.c.) \nn && + V \sum_{i} (b_{i}f_{i\sigma}^{\dagger}c_{i\sigma} +
H.c.) + \sum_{i}b_{i}^{\dagger} \partial_{\tau} b_{i} \nn && +
\sum_{i}f_{i\sigma}^{\dagger}(\partial_{\tau} +
\epsilon_{f})f_{i\sigma} + J \sum_{\langle ij \rangle} (
f_{i\sigma}^{\dagger}\chi_{ij}f_{j\sigma} + H.c.) \nn && + i
\sum_{i} \lambda_{i} (b_{i}^{\dagger}b_{i} + f_{i\sigma}^{\dagger}
f_{i\sigma} - 1) + NJ \sum_{\langle ij \rangle} |\chi_{ij}|^{2} .
\eqa Here, $c_{i\sigma}$ and $d_{i\sigma} =
b_{i}^{\dagger}f_{i\sigma}$ are conduction electron with a
chemical potential $\mu$ and localized electron with an energy
level $\epsilon_{f}$, where $b_{i}$ and $f_{i\sigma}$ are called
holon and spinon, representing hybridization and spin
fluctuations, respectively. The spin-exchange term for the
localized orbital is introduced as a competitor of  the
hybridization term, and decomposed via exchange hopping processes
of spinons, where $\chi_{ij}$ is a hopping parameter for the
decomposition. $\lambda_{i}$ is a Lagrange multiplier field to
impose the single occupancy constraint $b_{i}^{\dagger}b_{i} +
f_{i\sigma}^{\dagger} f_{i\sigma} = N/2$, where $N$ is the number
of fermion flavors with $\sigma = 1, ..., N$.

Performing the saddle-point approximation of $b_{i} \rightarrow
b$, $\chi_{ij} \rightarrow \chi$, and $i\lambda_{i} \rightarrow
\lambda$, one finds an orbital selective Mott transition as Kondo
breakdown at $J \approx T_{K}$. For $\langle b_{i} \rangle = 0$, a spin-liquid Mott insulator
 arises with a small area of the
Fermi surface in $J > T_{K}$ while for $\langle
b_{i} \rangle \not= 0$ a heavy Fermi liquid obtains with a large Fermi surface in
$T_{K} > J$ \cite{KB_Senthil,KB_Indranil,KB_Pepin}. Here, $T_{K} =
D \exp\Bigl(\frac{\epsilon_{f}}{N \rho_{c}V^{2}}\Bigr)$ is the
single-ion Kondo temperature, where $\rho_{c} \approx (2D)^{-1}$
is the density of states for conduction electrons with the half
bandwidth $D$. Reconstruction of the Fermi surface occurs at $J \simeq
T_{K}$.

The fluctuation-corrections are taken into account in the
Eliashberg framework \cite{KB_Indranil,KB_Pepin}. The main physics
is that the Kondo breakdown QCP is multi-scale. The dynamics of
the hybridization fluctuations is described by $z = 3$ critical
theory due to Landau damping of electron-spinon polarization above
an intrinsic energy scale $E^{*}$, while by $z = 2$ dilute Bose
gas model below $E^{*}$. The energy scale $E^{*}$ originates from
the mismatch of the Fermi surfaces of the conduction electrons and
spinons, one of the central aspects in the Kondo breakdown
scenario. Physically, one may understand that quantum fluctuations
of the Fermi surface reconfiguration start to be frozen at $T
\approx E^{*}$, thus the conduction electron's Fermi surface
dynamically decouples from the spinon's one below $E^{*}$. We show
that the Seebeck coefficient collapses at $E^{*}$, associated with
the Fermi surface reconstruction.

The thermoelectric power can be deduced from the following
transport equations \bqa && J_{el}^{c} = \sigma_{c}(E -
\nabla\mu_{c}) - p_{c} \nabla{T}  , \nn && J_{th}^{c} = T p_{c}(E
- \nabla\mu_{c}) - \kappa_{c} \nabla{T} , \\ && J_{el}^{f} =
\sigma_{f}(\epsilon - \nabla\mu_{f})- T p_{f} \nabla{T} , \nn &&
J_{th}^{f} = T p_{f} (\epsilon - \nabla\mu_{f}) - \kappa_{f}
\nabla{T} , \\ && J_{el}^{b} = \sigma_{b}(- E + \epsilon -
\nabla\mu_{b}) - p_{b} \nabla{T} , \nn && J_{th}^{b} = T p_{b} (-
E + \epsilon - \nabla\mu_{b}) - \kappa_{b} \nabla{T} , \eqa
defining three transport coefficients of $\sigma$, $p$, and
$\kappa$, where three species of conduction electrons, holons, and
spinons are taken into account for each response function, denoted
by $c$, $b$, and $f$, respectively. $J_{el(th)}$ is the electric
(thermal) current, and ${E}$, ${\epsilon}$, $\mu$, $T$ are an
external electric field, internal one, chemical potential and
temperature, respectively. The internal electric field  $\epsilon $
emulates the  ``back-flow'' condition ${J}_{el}^{f} +
{J}_{el}^{b} = 0$ \cite{Backflow}, central to transport in gauge theories. Taking $\mu_{c} = \mu_{f} -
\mu_{b}$ with an open-circuit boundary condition ${J}_{el}^{c} -
{J}_{el}^{b} = 0$, we find the total thermopower conductivity
$p_{t}(T)$ and electrical one $\sigma_{t}(T)$ \bqa && p_{t}(T) =
p_{c}(T) + \frac{ \sigma_{b}(T) p_{f}(T) - \sigma_{f}(T) p_{b}(T)
}{\sigma_{b}(T) + \sigma_{f}(T) } , \nn && \sigma_{t}(T) =
\sigma_{c}(T) + \frac{\sigma_{b}(T)\sigma_{f}(T)}{\sigma_{b}(T) +
\sigma_{f}(T)} . \eqa Then, the physical Seebeck coefficient is
given by $S_{t}(T) \equiv - p_{t}(T) / \sigma_{t}(T)$.

Around the QCP,  the Seebeck coefficient can be  simplified as
follows \bqa S_{HF}(T) \approx - \frac{p_{c}(T) +
p_{f}(T)}{\sigma_{c}(T) + \sigma_{f}(T)} , ~~~ S_{QC,SL}(T)
\approx - \frac{p_{c}(T)}{\sigma_{c}(T)}  ,  \eqa based on the
fact that $\sigma_{f}(T) \ll \sigma_{b}(T) \rightarrow \infty$ in
the HF phase and $\sigma_{f}(T) \gg \sigma_{b}(T) \rightarrow 0$
in both the quantum critical regime and spin liquid state.

Evaluating the thermopower conductivity in the one-loop level \bqa
&& p_{c(f)}(T) = N \rho_{c(f)} v_{F}^{c(f)2}
\int_{-\infty}^{\infty} d \omega \Bigl( - \frac{\partial
f(\omega)}{\partial \omega} \Bigr) \Bigl(\frac{\omega}{T}\Bigr)
\nn && \int_{-\infty}^{\infty} d k [\Im
G_{c(f)}(k,\omega+i\delta)]^{2} , \nonumber \label {eqn7} \eqa where
$G_{c(f)}(k,\omega+i\delta)$ is the full Green's function in the
Eliashberg approximation and $v_{F}^{c(f)}$ is the Fermi velocity
of conduction electrons (spinons), we find \be \begin{array}{l} \displaystyle{p_{c(f)}(T) \approx  -
\frac{\pi^{2}}{3}
\frac{2}{v_{F}^{c(f)2}Z_{c(f)}^{3}(k_{F}^{c(f)})} \Bigl(
\frac{\partial Z_{c(f)}(k)}{\partial k} \Bigr)_{k_{F}^{c(f)}} } \\
\displaystyle{\times  \frac{T}{Z_{c(f)}(T)} \sigma_{c(f)}(T) }\\
\\
\mbox{   where  } Z_{c(f)}^{-1}(T) = \left . 1 - \frac{\partial}{\partial \omega} \Re
\Sigma_{c(f)}(k_{F}^{c(f)},\omega) \right |_{\omega=T} , \\
\\
Z_{c(f)}(k) = 1 +
\frac{1}{v_{F}^{c(f)}} \frac{\partial}{\partial k} \Re
\Sigma_{c(f)} (k,E_{k}^{c(f)}) \end{array} \eq are wave-function
and dispersion renormalization \cite{Mahan_Comment}, respectively,
and $\Re \Sigma_{c(f)}(k_{F}^{c(f)},\omega)$ is the real part of
the electron (spinon) self-energy. This expression is basically
the same as the standard representation
\cite{Sign_Seebeck,pruschke,zlatic}, where the derivative of the
density of states with respect to frequency is replaced with that
of the dispersion renormalization function with respect to
momentum. Thus, the sign of the Seebeck coefficient is given to be negative when the Kondo resonance lies
below the Fermi energy, exactly the Yb case.

%\bqa && \Re \Sigma_{c}(\omega ) = \frac{1}{4\pi^{3}}
%\int_{0}^{\infty} d q q^{2} \int_{-1}^{1} d z
%\int_{-\infty}^{\infty} \frac{d \Omega }{\pi} \Im
%\chi_{b}(q,\Omega) \nn && \mathbf{P} \Bigl\{ \frac{ \Theta(-
%\alpha v_{F}^{c} q^{*} - \alpha v_{F}^{c} q z) -\Theta(-\Omega )
%}{\alpha v_{F}^{c} q^{*} + \alpha v_{F}^{c} q z + \omega - \Omega}
%\Bigr\} \eqa

An important quantity is the quasiparticle weight
$Z_{c(f)}(\omega)$, resulting from the linear dependence of
frequency in the thermopower expression. A singular logarithmic
temperature dependence is revealed in the quantum critical regime,
typical of the $z = 3$ quantum criticality in three dimensions
beyond the SDW theory \cite{indranil}.
%This logarithmic behavior should
%not be confused with the log correction in the electrical
%conductivity because the origin is different from each other,
%called the thermal log instead of the transport log
%\cite{indranil}.

The self-energy correction can be found within  the Eliashberg
theory, where quantum corrections are self-consistently introduced
in the one-loop level but vertex corrections are neglected
\cite{KB_Indranil,KB_Pepin}.
The dispersion renormalization function is obtained from the
renormalized band. In the HF phase the spinon and conduction
electron bands are hybridized. Resorting to these two
renormalization functions, we can provide a fit for the recent
experimental data for YbRh$_2$Si$_2$ \cite{hartmann}. The main
feature of the experimental results is that the Seebeck
coefficient divided by temperature is found to increase
logarithmically in temperature above a certain energy scale
$T_{M}$ and drops down abruptly in both the quantum critical
regime and antiferromagnetic phase while it saturates to a
constant in the heavy Fermi liquid. The Kondo breakdown QCP
scenario reproduces this main feature almost exactly as shown in
Fig. 1. The log$T$ divergence is explained by the $z = 3$ quantum
critical regime of the theory. In particular, the collapse at
$T_{M}$ is attributed to suppression of quantum fluctuations of
Fermi surfaces at $E^{*}$, thus $T_{M} = E^{*}$ in the Kondo
breakdown theory, where localized f-electrons cannot participate
in carrying entropy due to the Fermi surface decoupling.  In this
regime the question whether $S/T$ changes sign at very low
temperatures is still experimentally under scrutiny.
 In the Kondo  breakdown theory, the negative sign of the Seebeck coefficient
 is attributed to  the  presence of the fluctuating hybridization between the f-spinons
 and the conduction electrons, yielding to the formation of the Kondo resonance
 below the Fermi level. Below $E^*$ however, the spinon Fermi surface decouples
 from the conduction electron one, and $S/T$
 saturates to the value determined solely by the light conduction electrons.
 In the heavy Fermi liquid the Kondo
breakdown theory shows the saturation associated with Fermi liquid physics, but without changing sign.

To understand better the quantum critical region, we approximate
the self-energy and obtain the typical $z = 3$ form analytically
in three dimensions, \bqa \Sigma_{c(f)}(k_{F}^{c(f)},\omega >
E^{*}) = - \frac{m_{b} V^{2}}{6 v_{F}^{f(c)}} \omega \ln \Bigl(
\frac{\alpha D}{\omega} \Bigr) + i \frac{m_{b} V^{2}}{12 \pi
v_{F}^{f(c)}} |\omega| , \nonumber \eqa where $D$ is the
half-bandwidth and $m_{b} = (2 N V^{2} \rho_{c})^{-1}$ is the band
mass of holons. Then, the two renormalization functions are given
by \bqa && Z_{c}^{QC-1}(T) = 1 - \frac{m_{b} V^{2}}{6 v_{F}^{f}} +
\frac{m_{b} V^{2}}{6 v_{F}^{f}} \ln \Bigl( \frac{\alpha D}{T}
\Bigr) , \nn && \Bigl( \frac{\partial Z_{c}^{QC}(k)}{\partial k}
\Bigr)_{k_{F}^{c}} = \Bigl( \frac{\partial Z_{c}^{HF}(k;b
\rightarrow 0)}{\partial k} \Bigr)_{k_{F}^{c}} \approx -
\frac{1}{4\alpha q^{*}} , \eqa displaying the  $ \log
T$-dependence and negative sign due to the f-resonance below the
Fermi energy.

%%%%%%%%%%%%%%%%%%%%%%%%%%%%%%%%%%%%%%%%%%%%%%%%%%%%%%%%%%%%%%%%%%%%%%%%
%\begin{figure}[t]
%\vspace{4cm}
%\includegraphics[width=0.5\textwidth]{Seebeck.eps}
%\caption{ (Color online) Hybridization-fluctuation-induced AF QCP,
%where condensation probability $M_{s} \propto |\langle b_{n\sigma}
%\rangle|^{2}$ (blue) vanishes but antiferromagnetic correlations
%$\Delta$ (purple) still exist.}
%\label{fig1}
%\vspace{-0.5cm}
%\end{figure}
%%%%%%%%%%%%%%%%%%%%%%%%%%%%%%%%%%%%%%%%%%%%%%%%%%%%%%%%%%%%%%%%%%%%%%%%

%%%%%%%%%%%%%%%%%%%%%%%%%%%%%%%%%%%%%%%%%%%%%%%%%%%%%%%%%%%%%%%%%%%%%%%%
\begin{figure}[t]
\vspace{0.0cm}
\includegraphics[width=9.5 cm]{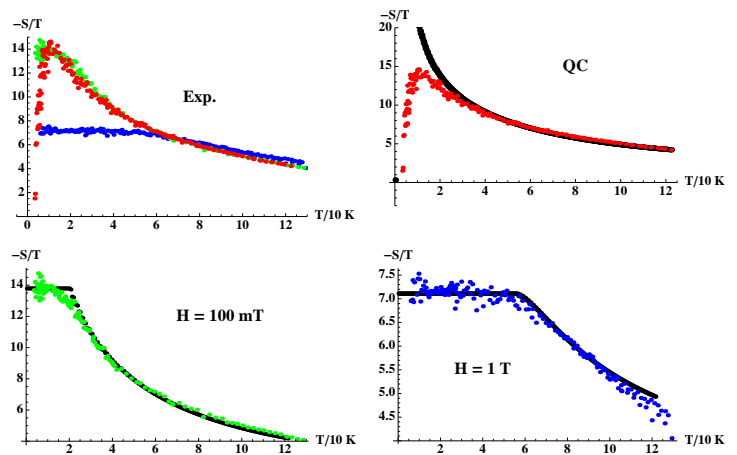} \caption{
(Color online) Fitting to the experimental data of Ref.
\cite{hartmann} based on the analytic expression of Eq.
(\ref{eqn7}). The values of the parameters are the half bandwidth
of the conduction electrons $D=990 K$ \cite{Band_Discussion}, of
the spinons $D_f=1 K$, and $E^* = 40 m K $. The upper left
quadrant represents the experimental data. The upper right
quadrant fits the QC regime. An overall adjusting coefficient $A=
0.33$ has been used. The holon mass is $ m= 6.0 \times 10^{-4} K$.
The lower left quadrant corresponds to a field of  $100 mT $. A
coefficient $A= 0.3$ has been used with  holon mass $m= 5.5 \times
10^{-4} K$. The lower right quadrant corresponds to a field of $1
T$. A coefficient $A= 0.0135$ has been used with holon mass $m=
-4.1 \times 10^{-4} K$. The  $ \log T $ behavior of the  $z=3$ QC
regime is correctly reproduced while the scale $E^*$ corresponding
to the reconfiguration of the Fermi surface is found to be a bit
small. } \label{fig1} \vspace{-0.5cm}
\end{figure}
%%%%%%%%%%%%%%%%%%%%%%%%%%%%%%%%%%%%%%%%%%%%%%%%%%%%%%%%%%%%%%%%%%%%%%%%

%%%%%%%%%%%%%%%%%%%%%%%%%%%%%%%%%%%%%%%%%%%%%%%%%%%%%%%%%%%%%%%%%%%%%%%%%
%\begin{figure}[t]
%\vspace{4cm}
%\includegraphics[width=0.5\textwidth]{Seebeck_Full_Log.eps} \caption{
%(Color online) } \label{fig3} \vspace{-0.5cm}
%\end{figure}
%%%%%%%%%%%%%%%%%%%%%%%%%%%%%%%%%%%%%%%%%%%%%%%%%%%%%%%%%%%%%%%%%%%%%%%%

%%%%%%%%%%%%%%%%%%%%%%%%%%%%%%%%%%%%%%%%%%%%%%%%%%%%%%%%%%%%%%%%%%%%%%%%
%\begin{figure}[t]
%\vspace{0.0cm}
%\includegraphics[width=0.5\textwidth]{Seebeck_Log.eps} \caption{
%(Color online) } \label{fig2} \vspace{-0.5cm}
%\end{figure}
%%%%%%%%%%%%%%%%%%%%%%%%%%%%%%%%%%%%%%%%%%%%%%%%%%%%%%%%%%%%%%%%%%%%%%%%

Lastly, we show that in the HF regime ($V>V_{c} $), the saturation
value of the Seebeck coefficient divided by the temperature
$\alpha\equiv - \frac{S_{HF}(V)}{T}$ is proportional to $B - B^*$,
where $B^*$ is  slightly displaced from the critical field $B_c$
of the antiferromagnetic QCP in YbRh$_2$Si$_2$. Resorting to the
condensation amplitude  $  b^{2} \approx \mathcal{C} \frac{
\ln\alpha^{-1}}{(1-\alpha)} \Bigl( \frac{-\epsilon_{f}}{D}\Bigr)
(J_{K} - J_{K}^{c})  $, where $\mathcal{C}$  is  a positive
numerical constant from the mean-field analysis, we obtain \bqa &&
\alpha \propto \mathcal{C} \frac{- \epsilon_{f} }{v_{F}^{f
2}q^{*2}} \frac{ \ln\alpha^{-1}}{(1-\alpha)} J_{K}^{c} ( J_{K} -
J_{K}^{R}) ,  \eqa where $J_{K}^{R} \equiv J_{K}^{c} - \frac{
v_{F}^{f 2}q^{*2}}{- \epsilon_{f} } \frac{(1-\alpha)}{
\ln\alpha^{-1}} \frac{1}{\mathcal{C} J_{K}^{c}} \approx
J_{K}^{c}$. Assuming $J_{K} \propto B$ valid near the QCP, we
conclude $\alpha \propto B - B^*$, consistent with the
experimental data of  Ref. \cite{hartmann}.

The weak point of the Kondo breakdown theory is the treatment of
the anti-ferromagnetism. However, the thermoelectric power is
insensitive against onset of anti-ferromagnetism as far as $E^{*}$
is larger than $T_{N}$, N\'eel temperature. Since localized
f-electrons cannot carry entropy when they are decoupled from
conduction electrons in the quantum level, the collapse should
occur from $E^{*}$ above $T_{N}$. Actually, this is observed in
YbRh$_2$Si$_2$, where $E^{*} \approx 100 mK$ and $T_{N} \approx 70
mK$ as the maximum N\'eel temperature \cite{hartmann}. This
collapse behavior is believed to survive beyond our approximation.

%On the other hand, the situation is somewhat complicated away from
%the QCP in the antiferromagnetic side, where $T_{N} > E^{*}$ is
%satisfied. In the two dimensional antiferromagnetic case out of
%the spin liquid phase, the whole "Fermi surface" of spinons can be
%gapped below the Neel temperature, and holon excitations become
%already suppressed even above $E^{*}$ because the holon
%self-energy given by the spinon-electron polarization displays gap
%physics. Then the abrupt drop in the Seebeck coefficient starts to
%occur from $T_{N}$ instead of $E^{*}$. In three dimensions Fermi
%surface exists generically, associated with cold regions. There
%will be no drastic measurable signatures at the N\'eel temperature
%for the Seebeck coefficient, and the abrupt drop occurs at $E^{*}$.

This feature differs from the SDW scenario completely, well
discussed in Ref. \cite{Pepin_KS}. Because quantum fluctuations
associated with the Fermi surface reconfiguration do not exist in
the SDW theory, the Seebeck coefficient should saturate to a
constant value below both the Neel temperature and HF coherence
one. It should display a symmetric configuration for $B < B_{c}$
and $B > B_{c}$.

Another important result of the Kondo breakdown theory applied to
YbRh$_2$Si$_2$ is that the sign of the Seebeck coefficient becomes
positive in the low temperature limit, changing from that in the
HF phase \cite{hartmann}. Since the spinon band is decoupled from
the conduction band, the Kondo resonance disappears even in the
dynamical level and contributions from only conduction electrons
give rise to a small positive constant, reflecting the physics of
the normal Fermi liquid. This  feature is difficult to interpret
within the SDW framework without the Fermi surface reconstruction.

One may ask why the Seebeck coefficient is sensitive to  $E^{*}$
while other quantities such as the specific heat coefficient
\cite{Kim_GR} and thermal conductivity \cite{Kim_TR} do not show
such serious dependence. The main difference between the
thermoelectric power and others is that the thermodynamic and
thermal transport coefficients can measure contributions of
neutral spinon excitations while the Seebeck coefficient does not.
This is the reason why only the thermopower is sensitive on the
Fermi surface reconstruction, argued to be a fingerprint for the
Kondo breakdown QCP.

Another important measurement is the Hall coefficient which also
reveals an interesting energy scale, referred as $T^{*}$ in the
heavy fermion side, below which the Hall coefficient $R_H$ shows
an abrupt decrease. This abrupt change has been attributed to the
proximity to a Kondo breakdown QCP \cite{Hall,Hall2}, whereas
other approaches have stressed the complexity of this measurement
and its sensitivity to minus fluctuations of the f-electron
chemical potential \cite{norman2}.
%
%An early study of this quantity for the Kondo breakdown theory
%predicts a jump in the Hall resistivity at the QCP
%\cite{coleman2}, but the issue of whether the Hall number changes
%abruptly or not at $T=0$  within this theory has not yet been
%addressed.
%
Since the Hall coefficient measures the Fermi surface
curvature, thus sensitive to the static formation of the heavy
fermion band, the Hall number shows its characteristic only in the
heavy fermion side while it does not in the localized side. On the
other hand, the thermoelectric power measures fluctuations of
Fermi surfaces, showing an interesting signature in the
antiferromagnetic side.
%
%Although the conduction band is decoupled
%from the localized one, quantum fluctuations between the
%hybridized (heavy fermion) band and decoupled one exist above
%$E^{*}$, resulting in the logarithmic increase of the Seebeck
%coefficient in the quantum critical region. Such Fermi surface
%fluctuations are frozen below $E^{*}$, thus the thermopower signal
%drops abruptly because only conduction electrons contribute.
%
We refer interested readers to our future publication for a
thorough study of the Hall effect within the Kondo breakdown
theory.

In this paper, we argue that the thermoelectric power is an
important measurement for quantum fluctuations of the Fermi
surface reconfiguration, which enables us to discriminate the
Kondo breakdown theory from the SDW scenario. It was demonstrated
to collapse at the energy scale associated with Fermi surface
fluctuations, identified with $E^{*}$ in the Kondo breakdown
theory.  We show here, that a careful fitting of the data can be
obtained within the Kondo breakdown theory for an energy scale of
the order of $40 mK$, a bit small compared to the experimental
value. The $z = 3$ quantum criticality of Fermi surface
fluctuations gives rise to the singular $ \log T$ behavior in the
quantum critical regime, consistent with the experimental data on
YbRh$_2$Si$_2$.

The local quantum critical scenario \cite{Si} does not have the
characteristic energy scale $E^{*}$ of the Kondo breakdown theory.
There, the mechanism for the abrupt collapse of the Seebeck
coefficient  is  not yet as clear as within the Kondo breakdown
theory. It would be very interesting to see how the two theories
of Kondo breakdown compare with respect to the fitting of the
experimental data.

We thank I. Paul for very helpful discussions and S. Hartmann,
N.Oeschler and F.Steglich for kindly providing their experimental
data to us. K.-S. Kim was supported by the National Research
Foundation of Korea (NRF) grant funded by the Korea government
(MEST) (No. 2010-0074542). CP acknowledges ICAM travel fellowship
and the Aspen Center for Physics where the idea of this work took
form.

\end{document}